\documentclass[epj,nopacs]{svjour}
\usepackage{cite}
\usepackage{epsf}
\usepackage{rotating}
\usepackage{here}

\def\kevc1{\ifmmode\mathrm{\ keV/{\mit c}}
           \else$\mathrm{\ keV/{\mit c}}$\fi}
\def\Mevc1{\ifmmode\mathrm{\ MeV/{\mit c}}
          \else$\mathrm{\ MeV/{\mit c}}$\fi}
\def\gevc1{\ifmmode\mathrm{ GeV/{\mit c}}
           \else$\mathrm{ GeV/{\mit c}}$\fi}
\def\kevc2{\ifmmode\mathrm{\ keV/{\mit c}^2}
          \else$\mathrm{\ keV/{\mit c}^2}$\fi}
\def\Mevc2{\ifmmode\mathrm{\ MeV/{\mit c}^2}
          \else$\mathrm{\ MeV/{\mit c}^2}$\fi}
\def\Gevc2{\ifmmode\mathrm{\ GeV/{\mit c}^2}
          \else$\mathrm{\ GeV/{\mit c}^2}$\fi}
\def\Gev2c2{\ifmmode\mathrm{\ GeV^2/{\mit c}^2}
          \else$\mathrm{\ GeV^2/{\mit c}^2}$\fi}

\def\pupp{\ifmmode{\mathrm{\ (GeV/\mit c)^2}}
          \else$\mathrm{\ (GeV/{\mit c})^2}$\fi}
\def\pum{\ifmmode{\mathrm{\ (GeV/c)^{-1}}}
          \else$\mathrm{\ (GeV/{\mit c})^{-1}}$\fi}
\def\pumm{\ifmmode{\mathrm{\ (GeV/c)^{-2}}}
          \else$\mathrm{\ (GeV/{\mit c})^{-2}}$\fi}

\def\pt  {$p_{t}$}
\def\xf  {$x_{F}$}

\def\Pp{\ifmmode{\mathrm p}
         \else${\mathrm p}$\fi}
\def\Pn{\ifmmode{\mathrm n}
         \else${\mathrm n}$\fi}

\newcommand{\dcy}{\mbox{$ \rightarrow $}}

\def\kmin {K$^{-}$}

\def\sigmamin {$\Sigma^{-}$}

 \def\PgS{\ifmmode\mathrm{\Sigma}
           \else$\mathrm{\Sigma}$\fi}
 \def\PgSm{\ifmmode\mathrm{\Sigma^-}
           \else$\mathrm{\Sigma^-}$\fi}
           
 \def\PgL{\ifmmode\mathrm{\Lambda}
          \else$\mathrm{\Lambda}$\fi}
 \def\PgLz{\ifmmode\mathrm{\Lambda^0}
          \else$\mathrm{\Lambda^0}$\fi}
 
 \def\PgX{\ifmmode\mathrm{\Xi}
           \else$\mathrm{\Xi}$\fi}
 \def\PgXm{\ifmmode\mathrm{\Xi^-}
           \else$\mathrm{\Xi^-}$\fi}
 \def\PgXz{\ifmmode\mathrm{\Xi^0}
           \else$\mathrm{\Xi^0}$\fi}
 \def\PgXmA{\ifmmode\mathrm{\Xi^-_{1320}}
           \else$\mathrm{\Xi^-_{1320}}$\fi}
 \def\PgXmB{\ifmmode\mathrm{\Xi^-_{1820}}
           \else$\mathrm{\Xi^-_{1820}}$\fi} 
 \def\PgXmC{\ifmmode\mathrm{\Xi^-_{1950}}
           \else$\mathrm{\Xi^-_{1950}}$\fi}          
 \def\PgXmD{\ifmmode\mathrm{\Xi^-_{1530}}
           \else$\mathrm{\Xi^-_{1530}}$\fi}
 \def\PgXmBC{\ifmmode\mathrm{\Xi^-_{1820,1950}}
           \else$\mathrm{\Xi^-_{1820,1950}}$\fi}     
 
 \def\PgXA{\ifmmode\mathrm{\Xi_{1320}}
           \else$\mathrm{\Xi_{1320}}$\fi}       
 \def\PgXB{\ifmmode\mathrm{\Xi_{1820}}
           \else$\mathrm{\Xi_{1820}}$\fi}             
 \def\PgXC{\ifmmode\mathrm{\Xi_{1950}}
           \else$\mathrm{\Xi_{1950}}$\fi}              
 \def\PgXD{\ifmmode\mathrm{\Xi_{1530}}
           \else$\mathrm{\Xi_{1530}}$\fi}               
 \def\PgXBC{\ifmmode\mathrm{\Xi_{1820,1950}}
           \else$\mathrm{\Xi_{1820,1950}}$\fi}             
               
 \def\PgXzA{\ifmmode\mathrm{\Xi^0_{1530}}
           \else$\mathrm{\Xi^0_{1530}}$\fi}
 \def\PgXzB{\ifmmode\mathrm{\Xi^0_{1690}}
           \else$\mathrm{\Xi^0_{1690}}$\fi}
 
 \def\PgXstar{\ifmmode\mathrm{\Xi^*}
           \else$\mathrm{\Xi^*}$\fi}             
 \def\PgXmstar{\ifmmode\mathrm{\Xi^{-*}}
           \else$\mathrm{\Xi^{-*}}$\fi}       
 
 \def\PgOm{\ifmmode\mathrm{\Omega^-}
           \else$\mathrm{\Omega^-}$\fi}
 \def\PK{\ifmmode\mathrm{K}
         \else$\mathrm{K}$\fi}
 \def\PKpm{\ifmmode\mathrm{K^{\pm}}
           \else$\mathrm{K^{\pm}}$\fi}
 \def\PKp{\ifmmode\mathrm{K^+}
          \else$\mathrm{K^+}$\fi}
 \def\PKm{\ifmmode\mathrm{K^-}
          \else$\mathrm{K^-}$\fi}
 \def\PKz{\ifmmode\mathrm{K^0}
          \else$\mathrm{K^0}$\fi}
 
 \def\Pgp{\ifmmode\mathrm{\pi}
          \else$\mathrm{\pi }$\fi}
 \def\Pgpm{\ifmmode\mathrm{\pi^-}
           \else$\mathrm{\pi^-}$\fi}
 \def\Pgpp{\ifmmode\mathrm{\pi^+}
           \else$\mathrm{\pi^+}$\fi}
 \def\Pgppm{\ifmmode\mathrm{\pi^{\pm }}
            \else$\mathrm{\pi^{\pm }}$\fi}
 \def\Pgpz{\ifmmode\mathrm{\pi^0}
           \else$\mathrm{\pi^0 }$\fi}
 
 \def\Pe{\ifmmode\mathrm{e}
         \else$\mathrm{e}$\fi}
 \def\Pep{\ifmmode\mathrm{e^+}
          \else$\mathrm{e^+}$\fi}
 \def\Pem{\ifmmode\mathrm{e^-}
          \else$\mathrm{e^-}$\fi}
 \def\Pgm{\ifmmode\mathrm{\mu}
          \else$\mathrm{\mu}$\fi}
 \def\Pgmm{\ifmmode\mathrm{\mu^-}
           \else$\mathrm{\mu^-}$\fi}
 \def\Pgmp{\ifmmode\mathrm{\mu^+}
           \else$\mathrm{\mu^+}$\fi}

\begin{document}

\hugehead 
\newcounter{str1}
\newcounter{str2}
\newcounter{str3}
\newcounter{str4}
\newcounter{str5}

\title{
Production of \PgXstar\ resonances in \PgSm\ induced reactions at 345 GeV/c} 
\titlerunning{~}
\authorrunning{~}
\subtitle{The WA89 Collaboration} 
\author{
\renewcommand{\thefootnote}{{\rm\alph{footnote}}}%
M.I.~Adamovich         \inst{8}  \and 
Yu.A.~Alexandrov       \inst{8}  \fnmsep
 \thanks{supported by the Deutsche Forschungsgemeinschaft,
           contract number436 RUS 113/465, and Russian Foundation for
           Basic Research under contract number RFFI 98-02-04096.}
                     \setcounter{str5}{\value{footnote}} 
 \and
S.P.~Baranov           \inst{8} \fnmsep \footnotemark[\value{str5}]  \and 
D.~Barberis            \inst{3}  \and 
M.~Beck                \inst{5}  \and 
C.~B\'erat             \inst{4}  \and 
W.~Beusch              \inst{2}  \and 
M.~Boss                \inst{6}  \and 
S.~Brons               \inst{5} \fnmsep \thanks{Now at TRIUMF, Vancouver, B.C., Canada V6T 2A3}  \and 
W.~Br\"uckner          \inst{5}  \and 
M.~Bu\'enerd           \inst{4}  \and 
C.~Busch               \inst{6}  \and 
C.~B\"uscher           \inst{5}  \and 
F.~Charignon           \inst{4}  \and 
J.~Chauvin             \inst{4}  \and 
E.A.~Chudakov          \inst{6} \fnmsep 
\thanks{Now at Thomas Jefferson Lab, Newport News, VA 23606, USA} 
                                \setcounter{str3}{\value{footnote}}\and
U.~Dersch              \inst{5}  \and F.~Dropmann            \inst{5}  \and J.~Engelfried          
\inst{6} \fnmsep \thanks{Now at Institudo de Fisica, Universidad San Luis Potosi, S.L.P. 
78240, Mexico} \and F.~Faller              \inst{6} \fnmsep \thanks{Now at Fraunhofer 
Institut f\"ur Solarenergiesysteme, D-79100 Freiburg, Germany} \and A.~Fournier            
\inst{4} \and S.G.~Gerassimov        \inst{5} \fnmsep \inst{8} \fnmsep \thanks{Now at 
Technische Universit\"at M\"unchen, Garching, Germany} 
                                \setcounter{str2}{\value{footnote}} \and
M.~Godbersen           \inst{5} \and 
P.~Grafstr\"om         \inst{2} \and 
Th.~Haller             \inst{5} \and 
M.~Heidrich            \inst{5} \and 
E.~Hubbard             \inst{5} \and 
R.B.~Hurst             \inst{3} \and 
K.~K\"onigsmann        \inst{5} \fnmsep 
\thanks{Now at Fakult\"at f\"ur Physik, Universit\"at Freiburg, Germany} 
                                \setcounter{str1}{\value{footnote}} \and
I.~Konorov             \inst{5} \fnmsep \inst{8} \fnmsep \footnotemark[\value{str2}] \and 
N.~Keller              \inst{6}  \and K.~Martens             \inst{6} \fnmsep \thanks{ 
Now at Department of Physics and Astronomy, SUNY at Stony Brook, NY 11794-3800, USA} \and 
Ph.~Martin             \inst{4} \and S.~Masciocchi          \inst{5} \fnmsep 
\thanks{Now at Max-Planck-Institut f\"ur Physik, M\"unchen, Germany} \and 
R.~Michaels            \inst{5} \fnmsep \footnotemark[\value{str3}]  \and U.~M\"uller            
\inst{7}  \and H.~Neeb                \inst{5}  \and D.~Newbold             \inst{1}  
\and C.~Newsom              \thanks{University of Iowa, Iowa City, IA 52242, USA} \and 
S.~Paul                \inst{5} \fnmsep \footnotemark[\value{str2}] \and J.~Pochodzalla         
\inst{5}  \and I.~Potashnikova        \inst{5}  \and B.~Povh                \inst{5}  
\and R.D.~Ransome             \thanks{Rutgers University, Piscataway, NJ 08854, USA.} 
\and Z.~Ren                 \inst{5}  \and M.~Rey-Campagnolle     \inst{4} \fnmsep 
\thanks{permanent address: CERN, CH-1211 Gen\`eve 23, Switzerland}  \and G.~Rosner              
\inst{7}  \and L.~Rossi               \inst{3}  \and H.~Rudolph             \inst{7}  
\and C.~Scheel              \thanks{NIKHEF, 1009 DB Amsterdam, The Netherlands}  \and 
L.~Schmitt             \inst{7} \fnmsep \footnotemark[\value{str2}] \and H.-W.~Siebert          
\inst{6}  \and A.~Simon               \inst{6} \fnmsep \footnotemark[\value{str1}] \and 
V.J.~Smith             \inst{1} \fnmsep \thanks{supported by the UK PPARC}  \and 
O.~Thilmann            \inst{6}  \and A.~Trombini            \inst{5}  \and E.~Vesin               
\inst{4}  \and B.~Volkemer            \inst{7}  \and K.~Vorwalter           \inst{5}  
\and Th.~Walcher            \inst{7}  \and G.~W\"alder            \inst{6}  \and 
R.~Werding             \inst{5}  \and E.~Wittmann            \inst{5}  \and 
M.V.~Zavertyaev        \inst{8}  \fnmsep \footnotemark[\value{str5}]  } 

\institute{ 
\renewcommand{\thefootnote}{{\rm\alph{footnote}}}%
             University of Bristol, Bristol, United Kingdom \and
             CERN, CH-1211 Gen\`eve 23, Switzerland. \and
             Genoa University/INFN, Dipt. di Fisica,I-16146 Genova, Italy. \and
             Grenoble ISN, F-38026 Grenoble, France.  \and
             Max-Planck-Institut f\"ur Kernphysik, Postfach
             103980, D-69029 Heidelberg, Germany. \and
             Universit\"at Heidelberg, Physikal. Inst., D-69120 Heidelberg Germany.
             \thanks{supported by the Bundesministerium f\"ur Bildung,
                     Wissenschaft, Forschung und Technologie,
                     Germany, under contract numbers 05~5HD15I, 06~HD524I and 06~MZ5265}
                     \setcounter{str4}{\value{footnote}} \and
             Universit\"at Mainz, Inst. f\"ur Kernphysik, D-55099 Mainz,
             Germany. \footnotemark[\value{str4}] \and
             Moscow Lebedev Physics Inst., RU-117924, Moscow, Russia.
              }
             
\abstract{ We report on a measurement of the differential and total cross sections of  
inclusive production of \PgXstar resonances in \PgSm\ - nucleus collisions at 345 \gevc1\ 
. The cross section for inclusive \PgXzA\ production is about a factor of 5 below that of 
\PgXmA\ hyperons. The products of cross section and branching ratio for the observed 
channels \PgXzB\ $\to$ \PgXm \Pgpp\ , \PgXmB\ $\to$ \PgXzA \Pgpm\ and  \PgXmC\ $\to$ 
\PgXzA\ \Pgpm\ are lower by yet another order of magnitude. The \PgXmB\ and \PgXmC\ 
resonances show significantly harder \xf\ and \pt\ distributions than \PgXm\ and \PgXzA\ 
hyperons. A comparison of the \xf -distribution to PYTHIA and QGSM predictions provides 
only a partial agreement.} 

\maketitle 

\section{Introduction}
\label{sec_1} The production of the ground state hyperons and also of their antiparticles 
has been studied extensively in previous experiments, using neutron, protons and \PgXm\ 
as beam particles with momenta ranging up to 600 \gevc1\ 
\cite{Bad72,Hun75,Bou79,Bou80,Bia81,Car85,Bia87,Bis86}. In a recent paper \cite{Ada97} we 
have reported cross sections for inclusive \PgXm\ production by \PgSm and \Pgpm of 
345~\gevc1 and by neutrons of 260~\gevc1\ mean beam momenta. These measurements confirmed 
the important role of the valence quark overlap between the beam projectile and the 
produced baryon at high \xf , the `leading particle effect'. 

Only few data exist on the production of excited hyperon resonances in such beams 
\cite{Bia81,Bia86,Sch90}. In ref. \cite{Sch90} a different \xf\ behaviour for \PgXz\ and 
\PgXzA\ was reported. This difference between the octet and the decuplet baryons was 
nicely reproduced in terms of the quark-diquark cascade model \cite{Tas88}. Within that 
model this feature is related to the diquark structure of the incident and produced 
baryons. 

Here we extend this study for the first time to orbitally excited baryons and present 
cross section measurements for {\PgXzA}, {\PgXzB}, {\PgXmB} and {\PgXmC} production by 
\PgSm\ of 345~\gevc1\ mean momentum. The results on the \xf -dependence of the cross 
sections and on the ratio of \PgXstar\ and \PgXm\ production will contribute to a better 
understanding of the complex process of hadron production in hadron beams. Furthermore, 
these data may help to explore the role of higher resonances for the spectral 
distributions of multiply strange hyperons which have recently been suggested 
\cite{Dum99} as an important diagnostic tool for the early freeze-out stage in 
ultra-relativistic heavy ion collisions. 

\section{Hyperon beam and experimental apparatus}
\label{sec_2} The hyperon beam was derived from an external proton beam of the CERN-SPS, 
hitting a hyperon production target placed 16~m upstream of the experimental target. 
Negative secondaries with a mean momentum of 345~\gevc1\ and a momentum spread  $\sigma 
(p) /p \approx 9\%$ were selected in a magnetic channel. The production angles relative 
to the proton beam were smaller than 0.5 mrad. At the experimental target, the beam 
consisted of \Pgpm, \kmin, \sigmamin\ and \PgXm\ in  the ratio 2.3 : 0.025 : 1 : 0.008. A 
transition radiation detector (TRD) made of 10 MWPCs interleaved with foam radiators 
allowed \Pgpm\-suppression at the trigger level. Typically, about $1.8\cdot 10^5 \ \PgSm$ 
and $4.5 \cdot 10^5 \ \Pgpm$ were delivered to the target during one SPS-spill, which had 
an effective length of about 1.5 s. More details can be found in \cite{Ale98}. 

The experimental target consisted of one copper and three carbon blocks arranged in a row 
along the beam, with thicknesses corresponding to 0.026 $\lambda_I$ and three times 
0.0083 $\lambda_I$, respectively. At the target, the beam had a width of 3~cm and a 
height of 1.7~cm. Microstrip detectors upstream and downstream of the target measured the 
tracks of the incoming beam particles and of the charged particles produced in the target 
blocks. The target was positioned 14~m upstream of the centre of the Omega spectrometer 
magnet \cite{Beu77} so that a free field decay region of 10~m length was provided for 
hyperon and $K_S$ decays. Tracks of charged particles were measured inside the magnet and 
in  the field-free regions upstream and downstream by MWPCs and drift chambers, with a 
total of 130 planes. The Omega magnet provided a field integral of 7.5 Tm, and the 
momentum resolution achieved was $\sigma(p) / p^2 \approx 10^{-4}$ \pum. 

Downstream of the spectrometer, a ring-imaging Che\-ren\-kov detector,
an electromagnetic calorimeter and a ha\-dron calorimeter
were placed. They were not used in the analysis presented here.

The main trigger selected about $25\%$ of all interactions, using multiplicities measured 
in microstrip counters upstream and downstream of the target, and in scintillator 
hodoscopes and MWPCs behind the Omega magnet. Correlations between hits in different 
detectors were used in the trigger to increase the fraction of events with high-momentum 
particles, thus reducing background from low-momentum decay-pions in the beam. In 
addition, a reduced sample of beam trigger events were recorded for calibration purposes. 
The results presented in this article are based on 100 million events recorded in 1993. 

\begin{figure}
\begin{center}
\leavevmode
\includegraphics[width=7cm]{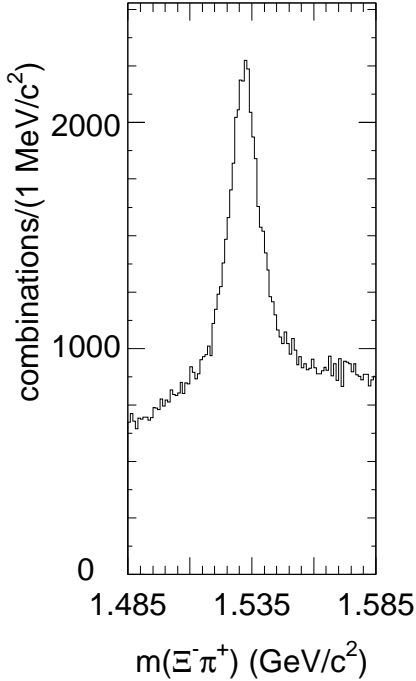}
\caption{ Effective mass distribution of $\Xi^- \pi^+$ combinations.}
\label{fig:ksy1520_mass}
\end{center}
\end{figure}

\begin{table}
\begin{center}
\begin{tabular}{|l|c|c|c|c|}   \hline
Particle    & Mass & Width &  $\sigma_{Data}$ & $\sigma_{MC}$ \\
\hline \PgL\dcy\Pp\Pgpm     &$ 1115.7\pm 0.1$ &   & 1.93  & 1.62 \\ \PgXm\dcy\PgL\Pgpm  
&$ 1321.0\pm 0.1$ &              & 2.68  & 2.32 \\ \PgOm\dcy\PgL\PKm    &$ 1672.5\pm 0.9$ 
&              & 2.4   & 2.1  \\ \PgXzA         &$ 1532.2\pm 0.5$ & $9.1      $  & 3.7   
& 3.2 \\ \PgXmB         & $1817 \pm 3$    & $23 \pm 13$  &       & 3.4  
\\ \PgXmC         & $1955 \pm 6$    & $68 \pm 22$  &       & 3.4  \\ \hline 
\end{tabular}
\end{center}
\caption{ Mean reconstructed masses and widths of
          \PgLz, \PgXm, \PgOm,
          \PgXzA, \PgXmB\ and \PgXmC\  and mass resolutions
           from data and Monte Carlo. All numbers are given in units of \Mevc2.}
\label{tab:mass_resol}
\end{table}

\section{Event selection}
\label{sec_3} In the following, we describe  the event selection for the decays \PgXzA\ 
$\rightarrow$ \PgXm \Pgpp\ and \PgXmstar $\rightarrow$ \PgXzA\Pgpm. Cuts imposed on the 
data sample in this analysis are very similar to those used in the measurement of the 
\PgXm\ production cross section \cite{Ada97} and in the search for \PgXzB $\to$ \PgXm 
\Pgpp\ decays \cite{Ada98}. 

First, \PgLz\ candidates were reconstructed from all pairs of positive and negative 
tracks which formed a vertex downstream of the target. The minimum distance of approach 
between the tracks was required to be smaller than 5 mm. For each candidate the effective 
\Pp \Pgpm\ mass and its error were calculated; the rms of the error was $\sigma_m 
\approx$ 2.2~\Mevc2. The reconstructed \Pp\Pgpm\ mass had to be within $\pm$15 \Mevc2 
around the known \PgLz\ mass \cite{PDG}, the error of the mass determination was required 
to be less than $5\cdot\sigma_m$. 

Next, \PgXm\ candidates were selected from all pairs of \PgLz\ candidates and negative 
tracks which formed a vertex downstream of the target. Again the minimum distance of 
approach between the reconstructed \PgLz\ track and the negative track was required to be 
smaller than 5 mm. A track corresponding to the \PgXm\ track as reconstructed from the 
decay cascade had to be found in the microstrip detectors downstream of the target. This 
track was used in the reconstruction of the production vertex (see below). The effective  
\PgL \Pgpm\ mass had to be within $\pm$15 \Mevc2\ of the \PgXm\ mass, for the mass 
resolution a value of $\leq 5\cdot\sigma_m$ was required, where $\sigma_m$ = 2.7 \Mevc2\ 
is the typical resolution. 

Then, we required a production vertex containing  at least one outgoing charged track in 
addition to the \PgXm\ track. The reconstructed vertex position had to be within a target 
block. In each coordinate an additional margin of 3$\cdot\sigma$ was allowed. The 
transverse distance between the \PgSm\ beam track  and the production vertex was required 
to be less than $6\cdot\sigma$ ($\sigma \approx 25 \mu m$). This requirement reduced the 
contributions from neutrons and \Pgpm\ originating from \PgSm\ decays upstream of the 
target. Events were also rejected if the beam track was connected to an outgoing track. 

Finally, \PgXzA\ candidates to be used in the search for \PgXmstar $\rightarrow$ \PgXzA 
\Pgpm\ decays were selected from all combinations of an accepted \PgXm\ with a positively 
charged particle emerging from the production vertex. The reconstructed \PgXm\Pgpp\ mass 
had to be within $\pm$10 \Mevc2\ and $\pm 10\cdot\sigma_m $ of the \PgXzA\ mass, 
$\sigma_m$ being 3.7 \Mevc2\ typically. 

\begin{figure}
\begin{center}
\leavevmode
\includegraphics[width=9cm]{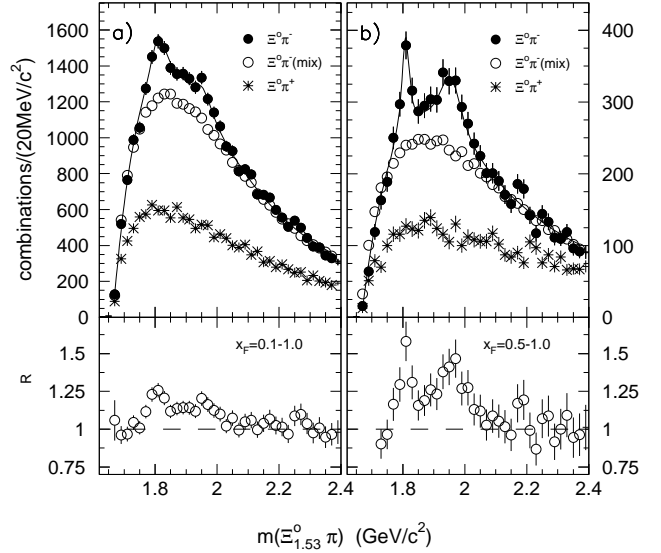}
\caption{ The \PgXm \Pgpp \Pgpm\ effective mass distribution in different \xf\ regions. 
The open circles show fake events generated by event mixing, the stars represent the 
background shape from "wrong sign" combinations. The lower parts display the ratio of the 
observed spectra and the backgrounds from event mixing.} \label{fig:ksy1820_mass} 
\end{center}
\end{figure}

\section{Observed {\PgXzA}, {\PgXmB} and {\PgXmC} signals.}
\label{sec_4} Fig.~\ref{fig:ksy1520_mass} shows the \PgXm\Pgpp\ mass distribution for all 
combinations of a \PgXm\ candidate with a positive particle  from the production vertex. 
A clear signal of {\PgXzA}\ decays is visible. In order to estimate the number of 
{\PgXzA}\ decays, the distribution was fitted by a combination of a Voigtian function 
(convolution of a Breit-Wigner and a Gauss distribution) with a Legendre polynomial of 
3rd order. The width of the underlying Breit-Wigner distribution was fixed to the known 
value $\Gamma(\PgXzA) = 9.1$ \Mevc2\ \cite{PDG}. The position of the signal $M = (1532.2 
\pm 0.5)$ \Mevc2\ is in good agreement with the known value of the \PgXzA\ mass,  $M = 
(1531.8 \pm 0.3)$ \Mevc2\ \cite{PDG}. The mass resolution resulting from the fit in this 
mass region is $\sigma_{Data} = 3.7$ \Mevc2, slightly higher than the value  $\sigma_{MC} 
= 3.2$ \Mevc2\ obtained from a Monte Carlo simulation. The same  15\% difference between 
measured and Monte Carlo mass resolutions was found for \PgLz, \PgXm\ and \PgOm\ decays 
(see Tab.~\ref{tab:mass_resol}). 

The \PgXm\Pgpp\Pgpm\ mass distribution for all combinations of a \PgXzA\ candidate with a 
negative particle from the interaction vertex is presented in 
Fig.\,{\ref{fig:ksy1820_mass}}. The left part of this figure shows the distribution for 
the range 0.1 $< x_F < $1. An  enhancement at 1820 \Mevc2\ is  visible. 
Fig.~\ref{fig:ksy1820_mass}b is for $x_F >$ 0.5. Here the signal at 1820 \Mevc2\ becomes 
much clearer and a second wider peak appears at about 1960 \Mevc2. 

The subtraction of the background under the  signals was done in the following way. We 
estimated the shape of the background distributions by ``event mixing'' \cite{Kop74}, 
combining the \PgXm\ from one event with all $\pi^+ \pi^-$ pairs from another event. The 
true and the mixed distributions were normalized to each other in the mass range of  
2.1-2.6~\Gevc2. The mixed distributions thus obtained are presented in the upper parts of 
Fig.~\ref{fig:ksy1820_mass} by open circles. In the lower panels of 
Fig.~\ref{fig:ksy1820_mass} we plot the ratios of the experimental mass distribution to 
this background distribution. Above 2100 \Mevc2\ the ratios are consistent with a 
constant value of one. The effective mass distributions of ``wrong'' sign combinations 
$\Xi^- \pi^+ \pi^+$ are indicated by the stars in the upper parts of 
Fig.~\ref{fig:ksy1820_mass}. They show also no structure in the region of interest and - 
except for a normalization constant - their shapes are consistent with the mixed 
distributions. 

The resulting mass and width of the {$\Xi^{-}_{1820}$}\  are $M = (1817 \pm 3)$ \Mevc2\ 
and $\Gamma = (23 \pm 13)$ \Mevc2, where the quoted errors are the statistical errors of 
the fit. The apparatus resolution used was $1.15 \cdot 3.4 \Mevc2 \approx 3.9$ \Mevc2\ 
(see Tab.~\ref{tab:mass_resol}). These values are in good agreement with the world 
average values $M = (1823 \pm 5)$ \Mevc2\ and $\Gamma = 24 ^{+15}_{-10}$ \Mevc2\ 
\cite{PDG}. 

The mass and width of the second peak are $M = (1955 \pm 6)$ \Mevc2\ and $\Gamma = (68 
\pm 22)$ \Mevc2, respectively. Also these values are in agreement with the average values 
from the earlier experiments, $M = (1950 \pm 15)$ \Mevc2\ and  $\Gamma = (60 \pm 20)$ 
\Mevc2\ \cite{PDG}. It is, however, not clear whether this signal has its origin in one 
or several states since several \PgXstar\ resonances were claimed to be observed in this 
mass region. In the following, we will treat the second peak as the decay of a single 
\PgXC\ state. 

\begin{figure}
\begin{center}
\leavevmode
\includegraphics[width=9cm]{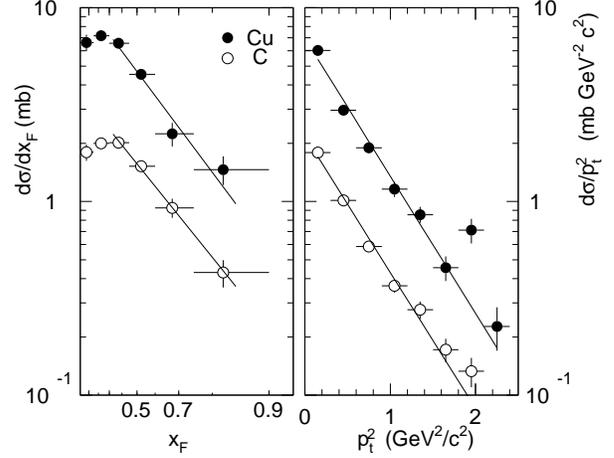}
\caption{ $d\sigma /dx_F$ [$\mu$b] and $d\sigma /dp^2_t$ [$\mu$b/\pupp] of inclusive 
\PgXzA\ production by \PgSm\ in copper and carbon.} \label{fig:ksy1530_xf_pt} 
\end{center}
\end{figure}

\begin{table}
\begin{center}
\begin{tabular}{|c|c|c|} \hline
$x_F$ & Copper & Carbon  \\ \hline 0.00 - 0.15&$  6.62 \pm  0.65 $ & $  1.80 \pm  0.17 $ 
\\ 0.15 - 0.30&$  7.18 \pm  0.36 $ & $  2.00 \pm  0.10 $ \\ 0.30 - 0.45&$  6.56 \pm  0.34 
$ & $  2.01 \pm  0.09 $ \\ 0.45 - 0.60&$  4.54 \pm  0.24 $ & $  1.52 \pm  0.08 $ \\ 0.60 
- 0.75&$  2.24 \pm  0.16 $ & $  0.93 \pm  0.05 $ \\ 0.75 - 0.90&$  1.46 \pm  0.12 $ & $  
0.43 \pm  0.03 $ \\ \hline 
\end{tabular}
\end{center}
\caption{ Differential cross section of \PgXzA\ production by \PgSm\ in copper and carbon 
as a function of \xf\ in mb.} \label{tab:ksy1520_diffxf} 
\end{table}

\begin{table}
\begin{center}
\begin{tabular}{|c|c|c|} \hline
$p_t^2$ & Copper & Carbon  \\ \hline 0.00 - 0.30&$  6.03 \pm  0.25 $ & $  1.79 \pm  0.07 
$ 
\\ 0.30 - 0.60&$  2.96 \pm  0.16 $ & $  1.02 \pm  0.05 $ \\ 0.60 - 0.90&$  1.90 \pm  0.13 
$ & $  0.59 \pm  0.04 $ \\ 0.90 - 1.20&$  1.16 \pm  0.10 $ & $  0.37 \pm  0.03 $ \\ 1.20 
- 1.50&$  0.86 \pm  0.08 $ & $  0.28 \pm  0.03 $ \\ 1.50 - 1.80&$  0.46 \pm  0.07 $ & $  
0.17 \pm  0.02 $ \\ 1.80 - 2.10&$  0.71 \pm  0.10 $ & $  0.13 \pm  0.02 $ \\ 2.10 - 
2.40&$  0.23 \pm  0.06 $ & $        -        $ \\ \hline 
\end{tabular}
\end{center}
\caption{ Differential cross section of \PgXzA\ production
           by \PgSm\
          in copper and carbon as a function of
          $p_t^2$ in  mb/(\gevc1)$^{2}$.}
\label{tab:ksy1520_diffpt}
\end{table}

\section{$\Xi^*$ production cross sections.}
\label{sec_5}
\subsection{The \PgXzA\ production cross section.}

The differential cross section as a function of the Feynman variable \xf\ and the squared 
transverse momentum $p_t^2$, were calculated using the following formula: 
\begin{equation}
      {  \sigma(x_F,p_t^2)  } =
            {{ 3/2 } \over { BR(\Lambda^0 \rightarrow p \pi^-)}} \cdot
            {{ N_{\Xi_{1530}} M} \over \varepsilon(x_F,p^2_t) \ N_{b}\
               \rho\ l\ N_A}.
\label{eq:crsec}
\end{equation}
Here $N_{\PgXD}$ is the number of observed \PgXD\ in the particular region of the 
corresponding kinematical variables and $\varepsilon$ denotes the overall acceptance 
including reconstruction and trigger efficiencies. $ N_{b}$ is the number of incoming 
beam particles tagged as \PgSm, corrected for beam contaminations (see below) and for 
losses due to the dead time of the trigger and the data acquisition system. $M$, $\rho$, 
$l$ are the atomic mass, the density and the length of the target, respectively; $N_A$ 
denotes Avogadro's number. The branching ratio of \PgXzA $\rightarrow $ \PgXm\ \Pgpp\  
decay is taken to be 2/3, as given by isospin conservation. 

The \PgXm\ contamination of the tagged beam was measured to be (1.26 $\pm$ 0.07)\% of  
the \PgSm\ flux \cite{Ada97}. The production cross section of the \PgXzA\ by \PgXm\ has 
been measured in a previous hyperon beam experiment \cite{Sch90} and the final values of 
our measurement were corrected for this contamination. Typically, this correction amounts 
to 6\%. 

About 12\% of the tagged beam particles  are fast  \Pgpm\ \cite{Ada97}. The cross section 
of \PgXzA\ production by pions has not been measured by any previous experiment. In 
contrast, the production cross section of \PgXm\ by pions has been measured in several 
experiments (see \cite{Ada97} and earlier references therein). From these measurements we 
derive a 2\% contribution from pions to the \PgXm\ production in our beam. Assuming the 
same ratio to holds for \PgXzA\ production, we subtracted this contribution from the 
final distributions. 

The tagged beam also contains a \PKm\ contamination of 2.1\%. The production cross 
section of \PgXzA\ by \PKm, however, is known only below 16~\gevc1\ \cite{Bau81}. It  
decreases slowly in the interval 4.2 - 16~\gevc1. Adopting the maximum value of $43~\mu 
b$ as the production cross section at 345~\gevc1\ we conclude that at most 0.4\% of the 
observed \PgXzA\ yield can be attributed to the \PKm\ content of the beam, which is 
negligible. 

The corrected differential production cross sections are shown in Fig.~ 
\ref{fig:ksy1530_xf_pt} and listed in tables \ref{tab:ksy1520_diffxf} and 
\ref{tab:ksy1520_diffpt} for copper and carbon targets, respectively. Only fit errors are 
quoted here. The cross sections were parameterized by a function of the form: 
\begin{equation}
\frac{d^2\sigma}{dp^2_t dx_F} = C (1 - x_F)^n \cdot exp (-bp_t^2),
\label{eq:param}
\end{equation}
which is based on quark counting rules and phase space arguments \cite{Bla74}. The three 
parameters $C$, $n$, and $b$ were assumed to be independent of $p_t^2$ and \xf. The 
values of $n$ and $b$ obtained from the fits are listed in Tab.~\ref{tab:ksy_crosst} for 
each target, and the fits are shown in the figures as straight lines over the fit range. 
No significant difference is observed between the values obtained from the copper and the 
carbon target. 

\begin{figure}
\begin{center}
\leavevmode
\includegraphics[width=9cm]{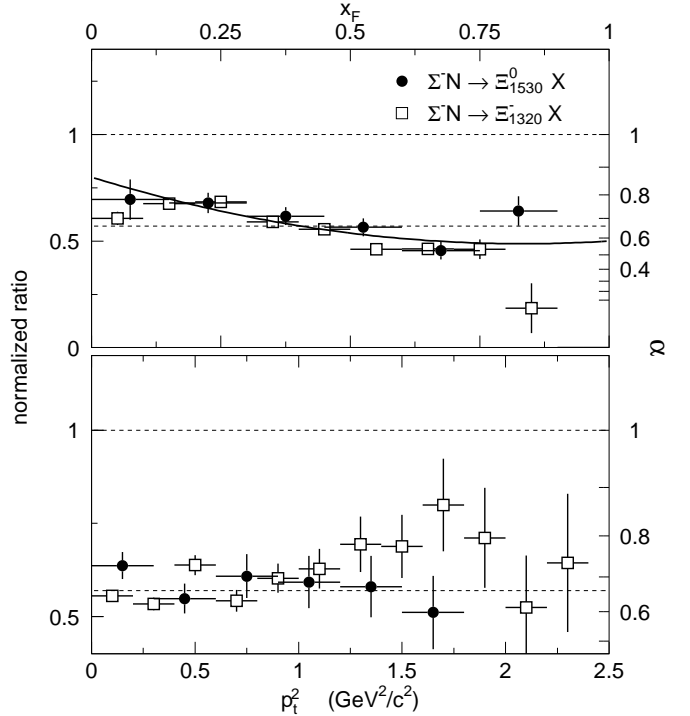}
\caption {Normalized ratio $R$ of the copper and carbon cross section (see Eqn. 
\protect\ref{eq:ratio}) for \PgXzA\ (dots) and \PgXmA\ (open squares, ref. \cite{Ada97}) 
production  by \PgSm\ as a function of \xf\ (top) and $p^2_t$ (bottom). The right scale 
indicates the exponent $\alpha$ in case of an $A^{\alpha}$ dependence. The solid line in 
the upper part marks a polynomial fit to a compilation of target attenuation factors 
given in ref. \protect\cite{Gei91}.} \label{fig:ksy1530_alfaxf} 
\end{center}
\end{figure}

Fig.~\ref{fig:ksy1530_alfaxf} shows the nuclear mass dependence of \PgXzA\ production as 
a function of \xf\ (top) and $p^2_t$ (bottom). Also shown are the values observed for 
\PgXm\ production \cite{Ada97}. The left-hand scales give the cross section ratio: 
\begin{equation}
{R=\frac{\sigma_{Cu}}{\sigma_C} \cdot \frac{A_C}{A_{Cu}}}
\label{eq:ratio}
\end{equation}
The right-hand scales shows the corresponding values of $\alpha$ in the conventional 
parametrization for the {\sl A} dependence: 
\begin{equation}
      \sigma = {\sigma}_0 \cdot A^{\alpha}
\end{equation}
The dashed lines correspond to $\alpha$=2/3 and $\alpha$=1.

The values for \PgXm\ and \PgXzA\ production are very similar to each other and to those 
observed in other hadroproduction processes which can be summarized as $\alpha (x_F) = 
0.8-0.75x_F+0.45x_F^2$ \cite{Gei91} (solid line in the upper part of 
Fig.~\ref{fig:ksy1530_alfaxf}). There is no visible dependence of $\alpha$ on $p^2_t$.

\begin{table}
\begin{center}
\begin{tabular}{|c|r|r|} \hline
\xf\ & \PgXmB\ & \PgXmC \\ \hline 0.10 - 0.25 & $ 30.6 \pm 10.8 $ & $      -     $ 
\\ 0.25 - 0.40 & $ 34.7 \pm  7.4 $ & $ 13.8 \pm  5.8  $ \\ 0.40 - 0.55 & $ 21.4 \pm  6.4 
$ & $ 4.6 \pm  5.2  $ \\ 0.55 - 0.70 & $ 15.4 \pm  5.1 $ & $ 19.9 \pm  5.8  $ \\ 0.70 - 
0.85 & $ 17.7 \pm  3.9 $ & $ 26.2 \pm  4.5 $ \\ 0.85 - 1.00 & $ 16.9 \pm  2.8 $ & $ 16.5 
\pm  2.8  $ \\ \hline 
\end{tabular}
\end{center}
\caption{ BR$\cdot d\sigma /dx_F$ per nucleon [$\mu$b] of inclusive
          \PgXstar\ production by \PgSm.}
\label{tab:ksy_diffxf}
\end{table}

\begin{table}
\begin{center}
\begin{tabular}{|c|r|r|} \hline
$p_t^2$ & \PgXmB\ & \PgXmC \\ \hline

0.00 - 0.30 & $ 16.7 \pm  4.9 $ & $ 8.1 \pm  4.6 $ \\
0.30 - 0.60 & $ 12.4 \pm  3.8 $ & $ 6.7 \pm  3.7 $ \\
0.60 - 0.90 & $ 14.2 \pm  3.6 $ & $ 10.9 \pm  3.1 $ \\
0.90 - 1.20 & $  9.6 \pm  2.9 $ & $ 8.6 \pm  2.5 $ \\
1.20 - 1.50 & $ 11.4 \pm  2.5 $ & $ 6.5 \pm  1.9 $ \\
1.50 - 1.80 & $  4.3 \pm  2.2 $ & $ 4.4 \pm  2.1 $ \\
1.80 - 2.10 & $  4.2 \pm  1.4 $ & $ 3.8 \pm  1.6 $ \\
2.10 - 2.40 & $  3.6 \pm  1.7 $ & $ 2.7 \pm  0.9 $ \\
\hline
\end{tabular}
\end{center}
\caption{ BR$\cdot d\sigma /dp^2_t$  per nucleon [$\mu$b/\pupp] of inclusive
          \PgXstar\ production by \PgSm.}
\label{tab:ksy_diffpt}
\end{table}

The total production cross sections per nucleus for copper and carbon are listed in 
Tab.~\ref{tab:ksy_crosst}. The quoted errors include an overall systematic error stemming 
from uncertainties of the efficiency determination (15\%), the trigger simulation (10\%) 
and the corrections for beam contaminations (7\%). Adding the systematic errors 
quadratically we estimated the total systematic error to be 20\% . 

The total production cross section per nucleon for \xf\ $>$ 0 was obtained by 
extrapolating the differential cross sections measured on Cu and C in each bin of \xf\ 
using the values of $\alpha$ obtained in the same bin. The result is $\sigma = (218 \pm 
44)~\mu b$, where the error is dominated by the systematic error quoted above. 

\begin{table*}
\begin{center}
\begin{tabular}{|l|r|c|c|c|c|} \hline
Particle   & \# events & $\sigma$ [mb]   & $\sigma$ [$\mu$b] & $n$ &    $b$         \\
  &           & per nucleus  & per nucleon  & ~ &(\gevc1)$^{-2}$ \\
\hline \PgXmA &                &   & $1000\pm40$ &               &              
\\ copper &              &$16.72 \pm 0.21$& & $1.97\pm0.04$ & $1.90\pm$0.04\\ carbon &              
& $5.40 \pm 0.07$& & $2.08\pm0.04$ & $2.00\pm$0.04\\ \hline \PgXzA\ &                &   
& $218\pm44$ &               &              \\ copper & $9333\pm171$ &$4.3 \pm 0.9 $ & & 
$1.32\pm0.08$ & $1.63\pm$0.06\\ carbon & $9547\pm157$ &$1.30 \pm 0.26$&  & $1.21\pm0.07$ 
& $1.62\pm$0.06\\ \hline 
  &  &  & $\sigma\cdot$BR & & \\ \hline
\PgXmB &$1045\pm116$& & $21\pm5$ & $ 0.2\pm 0.1 $ & $0.8\pm$0.20\\ \hline \PgXmC &$ 
872\pm137$& & $12\pm3$ & $-0.1\pm0.1$ & $0.7\pm$0.25\\ \hline 
\end{tabular}
\end{center}
\caption{ Number of reconstructed events and total inclusive \PgXstar\ production cross 
sections per nucleus and nucleon. $n$ and $b$ are the fit parameters of the differential 
cross sections with their fit errors (see text). The data for \PgXmA\ were taken from 
Ref. \cite{Ada97}.} \label{tab:ksy_crosst} 
\end{table*}

\begin{figure}
\begin{center}
\leavevmode
\includegraphics[width=9cm]{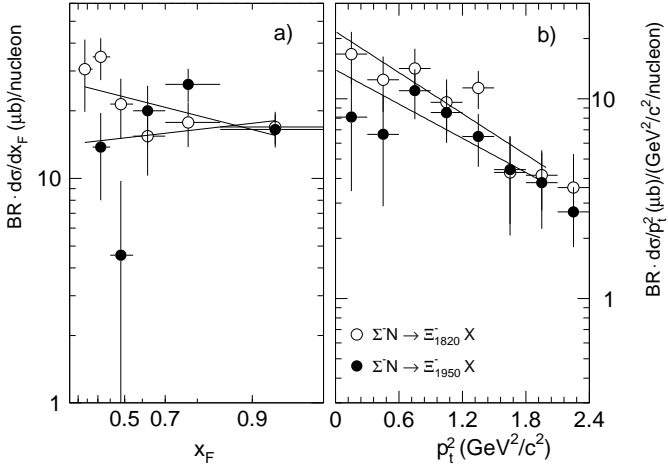}
\caption{ Differential cross sections times branching ratio per nucleon of inclusive 
\PgXmB\ and \PgXmC\ production  by \PgSm. The lines represent fits to the data according 
to Eqn. \ref{eq:param} } \label{fig:ksy1820_xf} 
\end{center}
\end{figure}

\subsection{The \PgXmB\ and \PgXmC\ production cross sections.}
The statistics of the observed \PgXmB\ and  \PgXmC\ signals is not sufficient to measure 
the cross section for each target separately. We therefore assume  that the inclusive 
production of these states has the same nuclear dependence as inclusive \PgXm\ and 
\PgXzA\ production, and analyze the data from the copper and carbon targets together. The 
differential cross sections are again parameterized by  Eqn.~\ref{eq:param}. 

Fig.~\ref{fig:ksy1820_xf} shows the differential cross section per nucleon multiplied by 
the branching ratio as a function of \xf\ and $p_t^2$. Only statistical errors are shown. 
The corresponding numbers are listed in tables~\ref{tab:ksy_diffxf} and 
\ref{tab:ksy_diffpt}. The inclusive production cross section per nucleon, $\sigma \cdot 
BR$ and the values of the fit parameters $n$ and $b$ are listed in 
Tab.~\ref{tab:ksy_crosst}. There are no experimental data on \PgXB\ or \PgXC\ production 
by \PgXm, \PKm\ or \Pgpm. Since, however, the correction to the total \PgXzA\ production 
cross section from the beam contaminations amounts to 7\% only, we neglect this 
correction here. A 7\% error from the uncertainties of this correction is included in the 
total systematic error of 20\%. 

\subsection{The \PgXzB\ production cross section.}
In a previous publication \cite{Ada98} we reported a measurement of $\sigma \cdot BR$ for 
\PgXzB $\to$ \PgXm \Pgpp\ production relative to \PgXzA\ production of
\[
\frac{\sigma \cdot BR (\PgXzB \rightarrow \PgXm \Pgpp) }
     {\sigma \cdot BR (\PgXzA \rightarrow \PgXm \Pgpp) } =
0.022 \pm 0.005.
\]
This value corresponds to the observed kinematic range,  0.1 $<$ \xf\ $<$ 1, and was 
extracted by assuming equal \xf\ and \pt\ dependence of  \PgXzB\ and \PgXzA\ production. 
With our present results we are now able to estimate a possible systematic variation of 
the production cross section by calculating the ratio of the detection efficiencies of 
the two states for three different \xf\ and $p_t^2$ dependencies of the \PgXzB\ 
production cross section, using the measured differential production spectra of the  
\PgXzA, \PgXmB\ and  \PgXmC. The  values of the parameters $n$ and $b$ used in the MC 
calculation were taken from Tab.~\ref{tab:ksy_crosst}. 

The resulting inclusive  production cross section multiplied by the branching ratio is  
shown in Tab.~\ref{tab:ksy1690_cross}. The quoted values are for the kinematical range 0 
$<$ \xf\ $<$ 1, the errors  are statistical only. The total systematic error on the scale 
of  the cross sections was estimated to be about 25\%. The values obtained range from 2.5 
to 6.8 $\mu$b. 

\section{Discussion}
\label{sec_6} In our experiment we observed four excited \PgX\ hyperons under the same 
conditions, which gives us a unique possibility to compare their production spectra. 

\begin{figure}
\begin{center}
\leavevmode
\includegraphics[width=7cm]{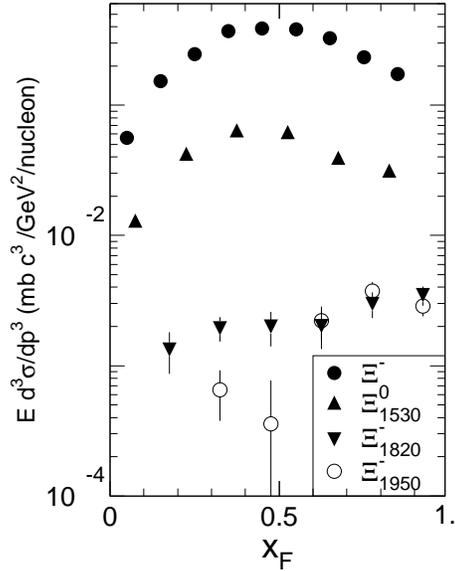}
\caption{ Invariant cross sections per nucleon of inclusive \PgXm\ and \PgXstar 
production by \PgSm.} \label{fig:ksy_invar} 
\end{center}
\end{figure}

In Fig.~\ref{fig:ksy_invar} we show the invariant cross sections for the observed signals 
including the ground state \PgXm \cite{Ada97}. For the \PgXmB\ and \PgXmC\, the values 
plotted are the cross sections multiplied with the branching ratio to \PgXzA \Pgpm. A 
strong leading particle effect is seen for the \PgX\ states produced. The comparison of 
the \xf\ behaviour of the cross sections shows that the octet and decuplet ground states 
\PgXm\ and \PgXzA\ are produced with  similar \xf\ spectra, with the \PgXzA\ cross 
section being smaller by a factor of about 5. The \xf\ spectra of the  excited states 
\PgXmB\ and \PgXmC\ are significantly harder (see Tab. \ref{tab:ksy_crosst}). This nicely 
illustrates that simple quark counting rules which reflect only the quark flavour 
contents of the involved baryons (see e.g. \cite{Bla74}) are not able to describe the 
hyperon production. 

The $p_t^2$ spectra of the \PgXzA\ shown on the right-hand side of 
Fig.~\ref{fig:ksy1530_xf_pt} have an exponential shape throughout the observed  
kinematical range 0 $< p^2_t <$ 2.4~\pupp, with slope parameters $b \approx$ 1.6~\pumm. 
The $p_t^2$ spectrum of the \PgXm\ has a similar slope with $b \approx$ 2.0~\pumm\ in the 
range  0 $< p^2_t <$ 1.0~\pupp, but a much harder $p_t^2$ spectrum above 1 \pupp\ 
corresponding to $b \approx$ 0.6 \pumm\ (see Fig.~5 of \cite{Ada97}). Also the  $p_t^2$ 
spectra of the  \PgXmB\ and \PgXmC\ shown in Fig.~\ref{fig:ksy1820_xf} are much harder 
than that of the \PgXzA\ or the soft part of the \PgXm\ spectrum. The  slope parameters 
$b$ = (0.8$\pm$0.2) \pumm\ for the \PgXmB\ and $b$ = (0.7$\pm$0.25) \pumm\ for the  
\PgXmC\ are close to the slope of the \PgXm\ distribution above 1 \pupp. 

Considering this agreement one may speculate whether decays of high lying \PgXstar\ 
resonances may contribute to the high-$p_t$ tail of the \PgXm\ hyperons. However, taking 
the decrease of the hyperon mass during the decay chain into account, the apparent slope 
parameter $b$ for the sequentially produced \PgXm\ will increase by typically a factor of 
$(1820/1320)^2 \approx 2$ to a value of about 1.5  \pumm\ . Furthermore, the two-step 
decays  
\begin{equation}
\PgXmBC \rightarrow \PgXzA \Pgpm \rightarrow \PgXm \Pgpp \Pgpm \label{Eqdecay1} 
\end{equation}
contribute not more than 3\% to the yield of \PgXm\ hyperons at large $p_t^2 
>$ 1~\pupp. The alternative decay chain 
\PgXmBC\ $\rightarrow $ \PgXmD \Pgpz $\rightarrow$ \PgXm \Pgpz \Pgpz\ increases this 
yield by an additional factor $5/4$. In Ref. \cite{Aps70} the ratio 
\begin{equation}
R= \frac{\PgXBC \rightarrow \PgX \Pgp + \PgX \Pgp \Pgp} {\PgXBC \rightarrow \PgXD \Pgp} 
\end{equation}
is estimated to be $1.5$  and $2.8$ for \PgXB\ and \PgXC\ decays, respectively. However, 
according to later observations \cite{Gay76} which are also consistent with earlier 
studies \cite{Ali69}, the contribution via non-resonant decays \PgXmB $\rightarrow$ 
\PgXm\Pgpz\ are a factor of 3 lower than the yield resulting from the resonant decay 
chain given in Eqn. \ref{Eqdecay1}. Therefore, we take the values from Ref. \cite{Aps70} 
as an upper limit and estimate that at most 10 \% of the observed \PgXm\ hyperons at high 
$p_t$ are decay products of negative \PgXmB\ or \PgXmC\ resonances. On the other hand, 
neglecting any non-resonant decays we obtain a lower limit of about 4\%. Of course 
additional higher lying \PgXmstar\ states and particularly neutral \PgX's will add 
further contributions. Nonetheless, both the relatively low yield and the expected 
increase of the slope parameter $b$ in sequential decays make it rather unlikely that 
sequential decays of high-lying \PgXstar\ resonances are the main origin of the 
high-$p_t$ tail of the \PgXm\ spectrum. 

\begin{table}
\begin{center}
\begin{tabular}{|c|c|} \hline
\xf\ spectrum & Cross section\\ as for     &  per nucleon [$\mu$b] \\ \hline \PgXzA\ & $ 
2.5 \pm 0.1 $     \\ \PgXmB\ & $ 5.6 \pm 0.2 $     \\ \PgXmC\ & $ 6.8 \pm 0.2 $     
\\ \hline 
\end{tabular}
\end{center}
\caption{ $\sigma \cdot$BR per nucleon of inclusive \PgXzB\
          production for different hypotheses about the \xf\ spectrum.}
\label{tab:ksy1690_cross}
\end{table}

The large background under the \PgXzB\ signal does not allow the \xf\ and \pt\ spectra to 
be measured. We expect the observed decay channel \PgXm\Pgpp\ to be the main decay 
channel, since the only other possible decay channels, $\PgL \overline{K}$ and 
\PgXzA\Pgpz, have thresholds of 1613 and 1667 \Mevc2\ respectively, and therefore a much 
smaller phase space than the \PgXm\Pgpp\ channel with its threshold at 1462~\Mevc2. Thus 
we expect the production cross section to be 2--10 $\mu b$ per nucleon in the range 0 $
<$ \xf\ $<$ 1, a factor of 20 or more below the corresponding \PgXzA\ cross section. 

There is only one publication about the branching ratio of \PgXmB $\to$ \PgXD \Pgp\ 
decay, giving a value 0.3 $\pm$ 0.15 \cite{Ali69}. This yields about (70 $\pm$ 35)~$\mu 
b$ per nucleon for inclusive \PgXmB\ production, which is about a factor of three below 
the \PgXzA\ cross section. Nothing is known about the \PgXmC\ decay branching ratios. The 
measured value of $\sigma \cdot BR = (12 \pm 3)~\mu b$ for \PgXmC\ production is smaller 
by a factor of 2 than the corresponding value  $\sigma \cdot BR = (21 \pm 5)~\mu b$ for 
\PgXmB\ production. 

\begin{figure}
\begin{center}
\leavevmode 
\includegraphics[width=9cm]{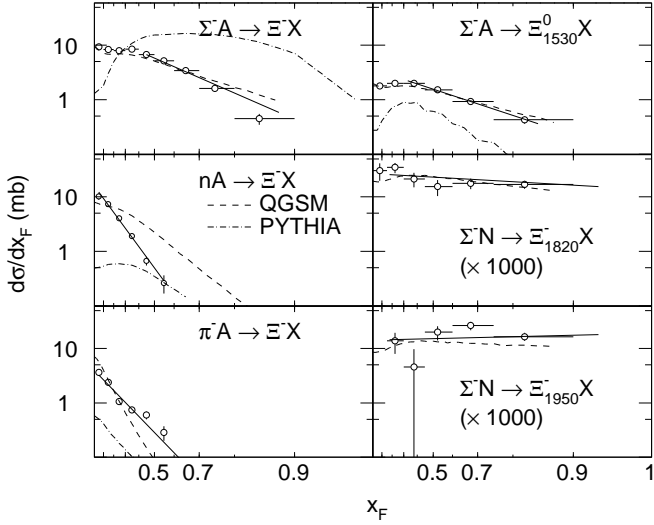}
\caption{ Comparison of differential cross sections of inclusive \PgX\ hyperons 
production with PYTHIA($- \cdot - \cdot$) and QGSM($---$) predictions. The QGSM cross 
sections for excited \PgX's were normalized to measured value. The solid lines represent 
the fit to the data according to Eqn. \ref{eq:param}.} \label{fig:ksy1320_xflog} 

\end{center}
\end{figure}

In Fig.~\ref{fig:ksy1320_xflog} we compare our experimental results with theoretical 
calculations based on PYTHIA \cite{PYTHIA} and the Quark-Gluon String Model (QGSM, see 
\cite{QGSM} and references therein). 

We have used PYTHIA with its default set of parameters, and made no effort to adjust them 
to better fit the experimental data (for reasons to be explained later). Among the 
essential details, we only mention the inclusion of elastic and diffraction processes 
(PYTHIA option {\tt MSEL=2}) and the usage of the Lund string fragmentation algorithm. 
The latter is closer to the QGSM ideology than `independent' or `cluster' fragmentation. 
The striking difference between the calculated cross sections, $\sigma_{\PgS \to \PgX}$ 
versus $\sigma_{\Pn \to \PgX}$ and $\sigma_{\Pgp \to \PgX}$, is due to the hierarchy of 
probabilities that PYTHIA attributes to the various species of quarks and diquarks 
appearing from the colour string fragmentation. The creation of a doubly strange diquark 
$ss$ (which is necessary to produce \PgX's from \Pgp\ or \Pn) is suppressed relative to 
the creation of a single strange quark (which is sufficient to form a \PgX\ from a \PgS) 
by a factor of 30. Such a suppression is also responsible for the relatively low 
abundance of \PgX\ at \xf\ $\simeq 0$ in the case of \PgS\ beams. 

Changing PYTHIA's default parameters in order to reach agreement with the data presented 
here, one would lose the consistency with other experimental results. For example 
reducing the suppression of doubly strange diquarks with respect to singly strange and 
non-strange diquarks one would no longer reproduce the cross section ratios observed in 
$\Xi/\Lambda/p$ inclusive production by protons \cite{Bou80}. A suppression of all 
diquarks with respect to single quarks would deteriorate the baryon to meson multiplicity 
ratios \cite{MARK,Blob}. Reducing the probability of single strange quark creation would 
cause difficulties with K/\Pgp\ and \PgL/\Pp\ ratios in the proton \cite{Blob} and pion 
\cite{Bre82} beams. 

The Quark-Gluon String Model offers two free parameters (for each isotopic family) that 
are the absolute probabilities to form the final state baryon via different fragmentation 
mechanisms. The baryon may be produced either directly from the beam diquark, or 
indirectly, in a remote part of the colour string (this includes quark, antiquark and 
antidiquark fragmentation cases). Although the absolute normalization of these two 
contributions is arbitrary, their \xf\ behaviour is strictly defined in the model. We 
have used the normalization freedom to tune the model to the \PgXA\ production in  \PgSm\ 
induced reactions on carbon. 

However, the QGSM predicts a stronger leading particle effect in the production of \PgX\ 
hyperons by neutrons than observed experimentally \cite{Ada97}. This discrepancy has its 
origin in the so called `baryon junction' which is introduced in the QGSM. The junction 
is considered to be as important for the baryon colour structure as the valence quarks 
themselves. The transmission of a junction from the projectile to a final state baryon 
creates a strong `leading particle effect' even for a neutron beam, though neutrons have 
only a single quark in common with \PgX\ hyperons. 

With respect to the production of excited hyperon resonances, we have extended the
original QGSM version by introducing a new assumption that the fragmentation into an 
excited state proceeds by an exchange of a shifted Regge trajectory (e.g., $\Delta$ or 
$N^*$) rather than by the ordinary nucleon trajectory. Then, the fragmentation functions 
acquire an extra factor, thus shifting in the same sense the \xf\ spectra. The absolute 
normalization of the production cross section still remains a free parameter that cannot 
be calculated within the theory. With the modification mentioned, the model appears to 
reproduce qualitatively the different shapes of the observed distributions. 

To summarize we have studied the production of several \PgXstar\ resonances in \PgSm\ - 
nucleus collisions at 345~\gevc1~. The cross section for inclusive \PgXzA\ production is 
about a factor of 5 below that of \PgXmA\ hyperons. The products of cross section and 
branching ratio for the observed channels \PgXzB\ $\to$ \PgXm \Pgpp, \PgXmB\ $\to$ \PgXzA 
\Pgpm and  \PgXmC\ $\to$ \PgXzA \Pgpm\ are lower by yet another order of magnitude. The 
\PgXmB\ and \PgXmC\ resonances show significantly harder \xf\ and $p_t$ distributions 
than \PgXm\ and \PgXzA\ hyperons. A comparison of the measured \xf\ distributions to 
PYTHIA and QGSM predictions provides only a partial agreement.

\section*{Acknowledgements}
It is a pleasure to thank J. Zimmer and G. Konorova for their support in setting up and 
running the experiment. We are also indebted to the staff of the CERN Omega spectrometer 
group for their help and support, to the CERN EBS group for their work on the hyperon 
beam line and to the CERN accelerator group for their continuous efforts to provide good 
and stable beam conditions. We thank S.~Brodsky, B.~Kopeliovich and O.~Piskounova for 
helpful discussions.

\end{document}